\documentclass[10pt,twocolumn,letterpaper]{article}

\usepackage{cvpr}              

\definecolor{cvprblue}{rgb}{0.21,0.49,0.74}
\usepackage[pagebackref,breaklinks,colorlinks,allcolors=cvprblue]{hyperref}
\usepackage{colortbl}
\usepackage{multirow}
\usepackage[T1]{fontenc}
\def\thickhline{\noalign{\hrule height.8pt}}
\definecolor{mygray}{gray}{.92}

\usepackage[accsupp]{axessibility}  

\def\paperID{2495} 
\def\confName{CVPR}
\def\confYear{2025}

\newcommand{\pub}[1]{{\color{gray}{\tiny{[{#1}]\!}}}}
\newcolumntype{x}[1]{>{\centering\arraybackslash}p{#1pt}}
\newcolumntype{y}[1]{>{\raggedright\arraybackslash}p{#1pt}}
\newcolumntype{z}[1]{>{\raggedleft\arraybackslash}p{#1pt}}

\newcommand{\eat}[2]{\setlength\tabcolsep{4pt}\renewcommand\arraystretch{1.10}
	\begin{tabular}{z{55}|y{33}}
		{#1} & {#2}
	\end{tabular}
}

\title{Silence is Golden: Leveraging Adversarial Examples to Nullify Audio Control in LDM-based Talking-Head Generation}

\author{Yuan Gan$^{1}$, \quad Jiaxu Miao$^{2}$\thanks{Corresponding author.}, \quad Yunze Wang$^{3}$,  \quad Yi Yang$^{1}$\\
$^1$ReLER, CCAI, Zhejiang University \\
$^2$School of Cyber Science and Technology, Sun Yat-sen University \\
$^3$Department of Statistics, University of Wisconsin–Madison
}

\begin{document}
\maketitle
\begin{abstract}
Advances in talking-head animation based on Latent Diffusion Models (LDM) enable the creation of highly realistic, synchronized videos. These fabricated videos are indistinguishable from real ones, increasing the risk of potential misuse for scams, political manipulation, and misinformation. Hence, addressing these ethical concerns has become a pressing issue in AI security. Recent proactive defense studies focused on countering LDM-based models by adding perturbations to portraits. However, these methods are ineffective at protecting reference portraits from advanced image-to-video animation. The limitations are twofold: 1) they fail to prevent images from being manipulated by audio signals, and 2) diffusion-based purification techniques can effectively eliminate protective perturbations. To address these challenges, we propose \textbf{Silencer}, a two-stage method designed to proactively protect the privacy of portraits. First, a nullifying loss is proposed to ignore audio control in talking-head generation. Second, we apply anti-purification loss in LDM to optimize the inverted latent feature to generate robust perturbations. Extensive experiments demonstrate the effectiveness of \textbf{Silencer} in proactively protecting portrait privacy. We hope this work will raise awareness among the AI security community regarding critical ethical issues related to talking-head generation techniques. Code: \href{https://github.com/yuangan/Silencer}{https://github.com/yuangan/Silencer}.
\end{abstract}    
\section{Introduction}
\label{sec:intro}



Talking-head animation~\cite{chen2019hierarchical, zhou2020makelttalk, guo2021ad, zhou2021pose, zhang2023sadtalker, wei2024aniportrait, gan2023efficient, tan2024edtalk} enables the creation of synchronized and highly realistic facial expressions based on audio and portrait images, producing videos that are often indistinguishable from authentic visual recordings. Recent advances in diffusion models~\cite{rombach2022high, hu2024animate, xu2024vasa, chen2024echomimic, cui2024hallo2} have markedly improved the realism of these animations. 
Consequently, this technological advancement increases the risks of misusing AI-Generated Content (AIGC) for scams, political manipulation, and misinformation. Mitigating these ethical risks has become a critical priority in AIGC security.

\begin{figure}
  \centering
  \includegraphics[width=0.475\textwidth]{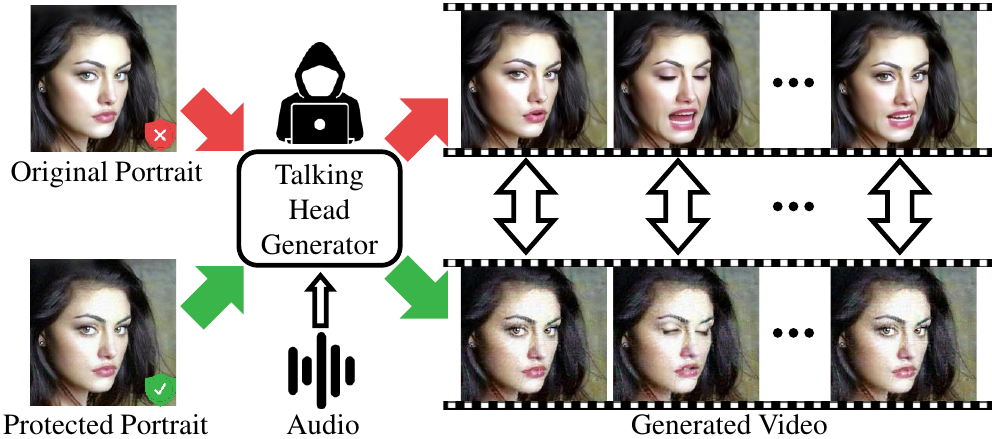}
  \vspace{-0.4cm}
  \caption{\textbf{Overview of Our Motivation.} Given an audio input, talking-head animation models can be exploited to generate fabricated videos using any portrait. To safeguard portrait privacy, we introduced \textbf{Silencer}, applying protective perturbations to ensure the portrait’s mouth remains closed in generated talking videos.}
  \label{fig:fig_1} 
  \vspace{-0.2cm}
\end{figure}

To address the ethical risks associated with AIGC, there are two primary approaches: passive defenses~\cite{rossler2019faceforensics++, dang2020detection, shiohara2022detecting,yan2023ucf, tan2024rethinking} and proactive defenses~\cite{li2023unganable, ruiz2020disrupting, liang2023adversarial, salman2023raising, liang2023mist, xue2023toward}. Passive defenses focus on detecting whether a video has been fabricated, making it useful for forensics. However, these approaches cannot prevent the infringement of personal privacy by deepfakes. When victims realize their privacy has been violated, the damage may already be irreparable. In contrast, 
proactive defenses offer superior protection by proactively shielding individuals from harm. These methods use adversarial perturbations on the input images to disrupt the outputs of the generative model.

Recent studies~\cite{liang2023adversarial, salman2023raising, liang2023mist, xue2023toward} explored proactive defenses against diffusion models, particularly those mimicry techniques based on Latent Diffusion Models (LDM). 
By adding perturbations to input images, these approaches have achieved copyright protection in diffusion-based mimicry. However, existing methods fail to protect privacy in the audio-driven talking-head generation, which utilizes LDM to animate the given portrait. Their limitations are twofold: {\bf\textit{1)}} they cannot prevent portraits from being animated by LDM-based talking-head models with a provided audio. Although these methods may reduce the quality of the generated video, this effect alone does not ensure privacy protection. {\bf\textit{2)}} diffusion-based purification techniques can remove these protective perturbations, counteracting the quality degradation and rendering the privacy measures ineffective.
Therefore, we need to generate robust adversarial perturbations that can nullify audio-driven facial movements and overcome purification techniques.



To address the above challenges with robust perturbations, we propose \textbf{Silencer}, a two-stage approach to proactively protect portrait privacy from animation by talking-head generation methods. In the first stage, we introduce a nullifying loss by disregarding audio control in the talking-head generation. Due to the lack of ground truth video, previous methods cannot be directly applied to talking-head generation. Our nullifying loss modifies the optimization objective of talking-head training to keep the portrait ``silent'', as shown in Fig.~\ref{fig:fig_1}. With the addition of adversarial noise via our nullifying loss, the generated talking videos tend to remain static, exhibiting low synchronization confidence. In the second stage, we design an anti-purification process using LDM to optimize the inverted latent feature, generating more robust perturbations. Since optimization in latent space does not have precise control over the outcomes in image space, directly applying nullifying loss to optimize the inverted latent feature would damage the adversarial portrait. Therefore, we use adversarial examples from the first stage to guide the optimization direction. To preserve the identity information, we apply a mask to the facial region halfway through the optimization process.



    
    
    


Overall, our main contributions are threefold:
\begin{itemize}
    
    \item We introduce a benchmark for assessing proactive protection measures against privacy threats posed by advanced LDM-based talking-head generation techniques.
    
    \item We propose \textbf{Silencer}, a two-stage paradigm, to proactively protect portrait privacy with robust adversarial perturbations. First, we introduce a nullifying loss that effectively renders a portrait “silent” in the talking-head generation. Second, we develop an anti-purification strategy to enhance the robustness of these perturbations against countermeasures.
    
    \item Extensive experiments are conducted to assess the efficacy of our Silencer. Our method achieves strong privacy protection with low synchronization confidence and exhibits resistance to purification-based attacks.

\end{itemize}

\section{Related Work}
\label{sec:RelatedWork}

\paragraph{Audio-Driven Talking-Head Animation.} Audio-driven talking-head generation has gained significant attention in recent years with the success of generative models~\cite{cao2023comprehensive, goodfellow2014generative,ho2020denoising,song2020denoising,tan2024navigation,rombach2022high, ruiz2023dreambooth, lipman2022flow}. Early methods~\cite{chung2017you, zhou2019talking, chen2019hierarchical, zhou2020makelttalk, guo2021ad, zhou2021pose, zhang2023sadtalker, wei2024aniportrait, gan2023efficient, tan2024edtalk} primarily relied on Generative Adversarial Networks (GANs)~\cite{goodfellow2014generative}. However, advancements in Latent Diffusion Models (LDMs) have led to more effective techniques~\cite{wei2024aniportrait, shen2023difftalk, stypulkowski2024diffused, tian2024emo, xu2024vasa, xu2024hallo, chen2024echomimic}. AniPortrait~\cite{wei2024aniportrait} improves visual quality and temporal consistency by projecting 3D representations as 2D landmarks in a diffusion model. In contrast, approaches like DiffTalk and Diffused Heads~\cite{shen2023difftalk, stypulkowski2024diffused} simplify the generation process by focusing on diffusion-based methods without relying on 3D models. Furthermore, EMO~\cite{tian2024emo} enhances expressiveness through a direct generation framework that eliminates the need for 3D models. VASA-1~\cite{xu2024vasa} performs efficient operations in the latent space for highly natural real-time generation. Hallo~\cite{xu2024hallo} incorporates cross-attention mechanisms and innovative audio-landmark training strategies to enhance generation quality and animation stability. In this paper, we adopt Hallo as the pre-trained talking-head model.

\paragraph{Adversarial Attacks in Diffusion Models.}
In the realm of adversarial attacks, early research introduced gradient-based methods that generate small perturbations to deceive neural network models~\cite{goodfellow2014explaining, kurakin2018adversarial, dong2018boosting, xie2019improving, dong2019evading, zhao2020towards, gao2020patch, long2022frequency}. Building on these methods, recent studies have applied adversarial attacks to diffusion models. AdvDM~\cite{liang2023adversarial} generates adversarial examples by optimizing latent variables during the reverse process of diffusion models. Photoguard~\cite{salman2023raising} "immunizes" images by adding imperceptible perturbations that prevent diffusion models from generating realistic manipulations. Extending the ideas of AdvDM and Photoguard, Mist~\cite{liang2023mist} incorporates semantic and texture loss designs to enhance cross-task transferability. Furthermore, Diff-Protect~\cite{xue2023toward} introduces Score Distillation Sampling (SDS) and highlights the encoder module as the main vulnerability affecting the robustness of diffusion models.

\paragraph{Purification and Anti-purification.}
Purification methods use generative models to remove adversarial noise before classification, thereby improving resistance to adversarial manipulations~\cite{samangouei2018defense, song2017pixeldefend, NEURIPS2019_378a063b, grathwohl2019your, hill2020stochastic, song2019generative, yoon2021adversarial, sandoval2023jpeg}. Building on this foundation, DiffPure~\cite{nie2022diffusion} utilizes the forward and reverse processes of diffusion models to purify adversarial examples. Moreover, GridPure~\cite{zhao2024can} introduces a grid-based iterative diffusion approach tailored to high-resolution images, enhancing purification effectiveness. PDM-Pure~\cite{xue2024pixel} uses pixel-space diffusion models as a universal purifier to mitigate adversarial noise.

To resist purification, ACA~\cite{chen2023contentbased} maps images onto a low-dimensional latent manifold of the generative model and optimizes adversarial objectives to enable diverse content generation and control. Additionally, DiffAttack~\cite{10716799} introduces an innovative, diffusion-based attack method to bypass existing purification defenses via latent feature optimization.

\section{Method}

\subsection{Preliminary}

\subsubsection{Audio-driven Talking-head Generation with LDM.}

Given a reference portrait $p$ and speech audio $a$, talking-head generation aims to generate realistic speaking videos synchronized with speech audio. To achieve this aim with powerful text-to-image LDM models, such as Stable Diffusion~\cite{rombach2022high}, recent works follow a common pipeline, which utilizes ReferenceNet and audio signals to guide the animation process. ReferenceNet has the same architecture as the LDM network, which extracts appearance features from reference images for guidance. As shown in Fig.~\ref{fig:fig_ldm}(b), talking-head LDM employs spatial attention to preserve intricate appearance features from the reference image. By integrating these appearance features, the model accurately captures the reference portraits, allowing for precise manipulation of facial expressions with audio inputs. Hence, during the training phase, an associated talking frame $f_i$ is encoded into a latent representation $z_0$ with the encoder of Variational AutoEncoder (VAE)~\cite{kingma2014autoencoding, esser2021taming}: $z_0 = \mathcal{E}(f_i)$. The diffusion process across $T$ timesteps then transforms this latent representation to a Gaussian noise $z_T\sim \mathcal{N}(0,1)$. The goal of the training is to progressively denoise $z_T$ to produce a realistic talking-head frame that not only preserves the visual characteristics of the reference portrait $p$ but also synchronizes the lip movements with the audio frame $a_i$. To achieve this aim, the training loss is defined by the following objective function:

\begin{figure}
  \centering
  \includegraphics[width=0.50\textwidth]{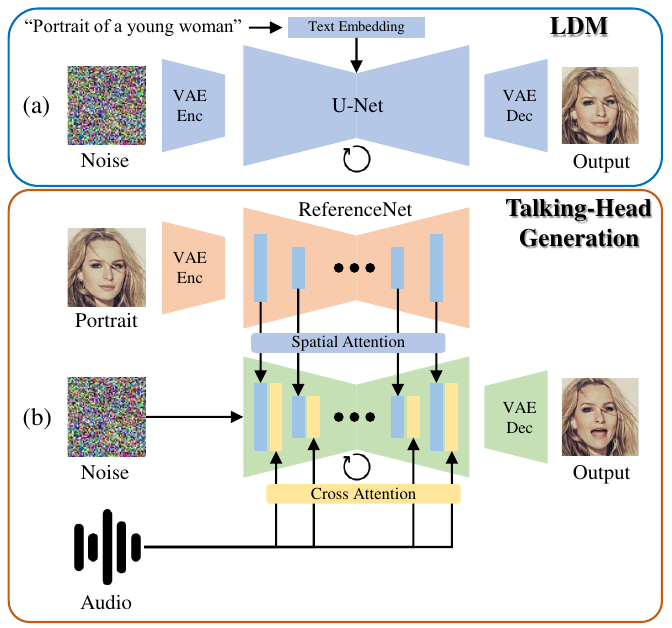}
  \vspace{-0.2cm}
  \caption{\textbf{LDM and LDM-based Talking-head Generation Framework.} (a) The inference process of latent diffusion models. Given random noise and text, LDM can generate a semantically coherent image through iterative denoising. (b) The talking-head generation framework. Given a portrait and audio frame, the talking-head generation model can produce a lip-sync video frame with realistic facial expressions. }
  \label{fig:fig_ldm} 
  \vspace{-0.3cm}
\end{figure}

\begin{equation}
\mathcal{L}_{ldm} = \mathbb{E}_{\mathcal{E}(f_i), p, a_i, \epsilon, t} \left[ \| \epsilon - \epsilon_\theta(z_t, t, p, a_i) \|^2_2 \right],
\label{eq:ldm_loss_ori}
\end{equation}

where $\epsilon \sim \mathcal{N}(0, 1)$ is a Gaussian noise, $\epsilon_\theta$ represents the denoising U-Net model that processes the noisy latent variable $z_t$ at each timestep $t$ along with the conditional inputs, $p$ is the reference portrait, $a_i$ is the $i$-th frame of the talking-head audio.

The ability to use any person's portrait as a reference in talking-head generation raises significant privacy concerns. To mitigate this, we propose a proactive defense mechanism centered around an open-source, advanced LDM-based talking-head generation method~\cite{xu2024hallo}.



\subsubsection{Adversarial Examples for LDM}
Adversarial examples can protect images from LDM-based mimicry by finding the appropriate perturbations that can effectively cause LDM models to generate visually corrupted outputs. Previous studies have used two objective functions to exploit vulnerabilities in the diffusion model: 

\begin{itemize}
    
    \item Semantic loss~\cite{liang2023adversarial} is the training loss of LDM, which disrupts the denoising process, directing the model to produce samples that differ from the real image:
    
    \begin{equation}
    \mathcal{L}_{S} = \mathbb{E}_{t, \epsilon} \mathbb{E}_{z_t} \|\epsilon - \epsilon_{\theta}(z_t, t) \|_2^2
    \end{equation}
    
    \item Texture loss~\cite{salman2023raising} attacks the VAE encoder $\mathcal{E}(\cdot)$ by steering the latent representation of the input image $x$ towards a target latent derived from another image $y$:
    \begin{equation}
    \mathcal{L}_{T} = -\| \mathcal{E}(x) - \mathcal{E}(y) \|_2^2
    \end{equation}
    
\end{itemize}

The final objective $\mathcal{L}_{adv}$ can be either semantic loss $\mathcal{L}_{S}$, texture loss $\mathcal{L}_{T}$, or both. Then PGD~\cite{madry2017towards} is chosen to generate adversarial examples with projected gradient ascent:

\begin{equation}
x^{n} = \mathcal{P}_{B_{\infty}(x, \delta)} \left[ x^{n-1} + \eta \, \text{sign} \, \nabla_{x^{n-1}} \mathcal{L}_{\text{adv}}(x^{n-1}) \right]
    \label{eq:PGD}
\end{equation}
where $x^n$ is the adversarial example at the $n$-th iteration, $\mathcal{P}_{B_{\infty}(x, \delta)}[\cdot]$ projects the adversarial output onto the $\ell_\infty$ ball centered at $x$ with budget $\delta$, $\eta$ is the step size.

\begin{figure*}
  \centering
  \includegraphics[width=1.\textwidth]{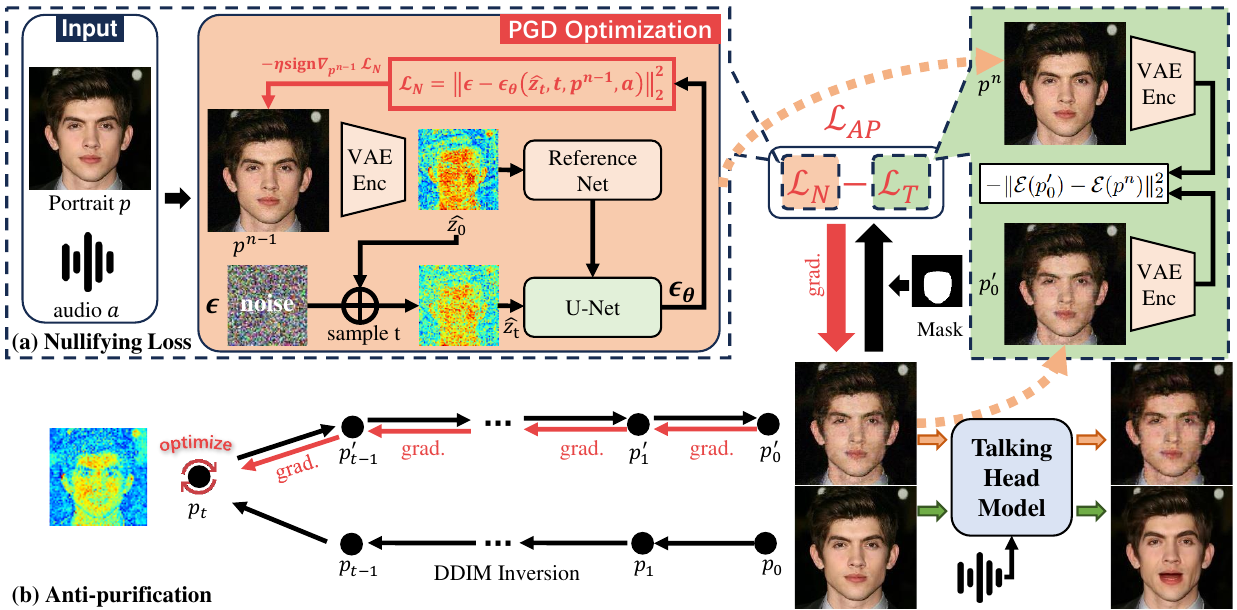}
  \vspace{-0.6cm}
  \caption{\textbf{Silencer Framework.} (a) In stage I, adversarial samples $p^n$ are generated through iterative PGD optimization using our proposed nullifying loss. These samples can avoid the influence of audio in the talking-head model. (b) In stage II, our anti-purification process is employed to optimize the inverted latent features, generating more robust perturbations capable of resisting purification. }
  \label{fig:fig_main} 
  \vspace{-0.3cm}
\end{figure*}

\subsection{Silencer}

To protect portrait privacy, a straightforward approach is to directly apply semantic loss~\cite{liang2023adversarial} to the talking-head animation task. However, this naive method presents two major challenges: First, unlike image generation tasks, we lack ground truth frames $f_i$, synchronized with audio frame $a_i$ for any arbitrary input portrait in talking-head animation. Second, we observed that noise-based perturbations introduced for privacy protection can be neutralized by purification methods. These purification methods counteract our protective measures, effectively compromising the intended privacy safeguards.
In the following sections, we propose \textbf{Silencer}, a two-stage method to address these challenges. 



\subsubsection{Silencer-I: Nullifying Loss}
To disrupt the denoising process and generate more artifacts in the edited images, semantic loss is used to optimize the perturbations by increasing or decreasing the LDM training loss. To calculate the training loss in Eq.~\ref{eq:ldm_loss_ori}, the ground truth frame $f_i$ and the corresponding latent representation $z_0$ are essential. A straightforward approach is to employ existing talking-head models to generate a synchronized video frame $f_i$. However, this has two drawbacks. {\bf\textit{1)}} Due to the time-consuming process of LDM inference, it is inefficient to generate fake ground truth under complex talking-head generation frameworks. {\bf\textit{2)}} Generating fake ground truth presents a paradox: protecting portrait privacy requires it to first be compromised. 

Unlike semantic loss, which requires ground truth, texture loss operates without this requirement, relying instead on a target image. Despite this advantage, texture loss does not directly affect the synchronization of the generated videos. This is primarily due to the changes of the LDM network architecture, as shown in Fig.~\ref{fig:fig_ldm}. Unlike traditional diffusion models, the LDM-based animation framework introduces conditions using ReferenceNet, which makes texture loss fail to eliminate the influence of the audio. Unless the face is fully obscured, the portrait can still be driven by audio. Hence, incorporating audio signals in the training of adversarial perturbations is crucial.

\begin{table*}
\setlength{\tabcolsep}{5pt}
\begin{center}
\small
\resizebox{0.995\textwidth}{!}{
\setlength\tabcolsep{6pt}
\renewcommand\arraystretch{1.}
\begin{tabular}{z{65}y{22}||c c c c | c c c c}
\hline \thickhline
\rowcolor{mygray}
 & &  & &\hspace{-4.2em}CelebA-HQ~\cite{karras2018progressive}     & & & &\hspace{-4.2em}TalkingHead-1KH~\cite{wang2021facevid2vid} & 
\\ \cline{3-10} \rowcolor{mygray} 
              \multicolumn{2}{c||}{\multirow{-2}{*}{Method}} & V-PSNR/SSIM$\downarrow$ &FID$\uparrow$ & Sync$\downarrow$ & M-LMD$\uparrow$ & V-PSNR/SSIM$\downarrow$ &FID$\uparrow$ & Sync$\downarrow$ & M-LMD$\uparrow$ \\ \hline \hline
AdvDM(+)~\cite{liang2023adversarial}\!\!\!&\!\!\!\pub{ICML23}\! & 17.95/\textcolor{blue}{0.4575} & 78.40 & 5.6150 & 2.0425& 19.09/\textcolor{red}{0.4437} & 178.92 & 3.8146 & 1.7581  \\
AdvDM(-)~\cite{xue2023toward}\!\!\!&\!\!\!\pub{ICLR24}\!  & \textcolor{red}{16.29}/0.4998 & 47.34 & 6.6670 & 2.1366& \textcolor{red}{17.42}/0.5556 & 52.99 & 5.2399 & 1.7244  \\
PhotoGuard~\cite{salman2023raising}\!\!\!&\!\!\!\pub{Arxiv23}\! & 17.67/0.4763 & 126.09 & 5.8875 & 2.0800& 18.76/0.5167 & \textcolor{blue}{186.84} & 3.3784 & \textcolor{red}{1.9023} \\
Mist~\cite{liang2023mist}\!\!\!&\!\!\!\pub{Arxiv23}\! & 17.80/0.4753 & \textcolor{blue}{134.44} & 5.9052 & 2.1173& 19.13/0.5241 & \textcolor{red}{221.58} & 3.0552 & 1.7787 \\
SDS(+)~\cite{xue2023toward}\!\!\!&\!\!\!\pub{ICLR24}\!   & 17.79/\textcolor{red}{0.4569} & 67.23 & 5.8668 & 2.1009& 19.11/\textcolor{blue}{0.4464} & 139.20 & 4.4844 & 1.6760 \\
SDS(-)~\cite{xue2023toward}\!\!\!&\!\!\!\pub{ICLR24}\!  & \textcolor{blue}{16.54}/0.4964 & 51.20 & 6.6743 & 2.0737& \textcolor{blue}{17.57}/0.5462 & 57.07 & 5.1954 & 1.7301  \\
SDTS(-)~\cite{xue2023toward}\!\!\!&\!\!\!\pub{ICLR24}\!  & 17.23/0.4828 & 89.70 & 6.4003 & 2.1024& 18.86/0.5496 & 139.87 & 3.8825 & \textcolor{blue}{1.8579}  \\ \hline
\eat{\textbf{Silencer}}{Stage I} &  & 19.02/0.5104& 124.07 & \textcolor{blue}{4.0644} & \textcolor{blue}{2.2008}& 20.61/0.5692& 168.85 & \textcolor{red}{1.7966} & 1.8025 \\
\eat{\textbf{Silencer}}{Stage II} & & 19.01/0.5111& \textcolor{red}{156.99}& \textcolor{red}{3.9685}& \textcolor{red}{2.2108}& 20.44/0.5718& 185.87& \textcolor{blue}{2.0017}& 1.8427
\\ 
\hline
\multicolumn{2}{c||}{Ground Truth} &  $\infty$ /1.00 &0.00     &6.4041 &0.0000 &  $\infty$  /1.00 &0.00     &5.4842  &0.0000 \\
\hline
\end{tabular}
}
\end{center}
\vspace{-.4cm}
\caption{\small{\textbf{Quantitative Comparisons with State-of-the-art Methods on CelebA-HQ~\cite{karras2018progressive} and TalkingHead-1KH~\cite{wang2021facevid2vid}.} "$\uparrow$": higher is better. "$\downarrow$": lower is better. \textcolor{red}{Red}: the 1st score. \textcolor{blue}{Blue}: the 2nd score.}}
\vspace{-.2cm}
\label{tab:tab_1}
\end{table*}


Given the limitations of existing loss functions regarding ground truth requirements and audio signal integration, we propose training adversarial perturbations that are both audio-aware and independent of ground truth. We observe that forcing the generated result to stay ``silent'' is an effective way to disrupt audio-visual synchronization, avoiding the need to directly attack the talking-head training process. To nullify the effect of audio signals $a$, we treat the reference portrait $p$ as the ground truth of the talking-head generation. Hence, we propose a nullifying loss to disrupt the audio-visual synchronization efficiently with the following formulation:

\begin{equation}
\mathcal{L}_{N} = \mathbb{E}_{t} \mathbb{E}_{\mathcal{E}(p), p, a_i, \epsilon} \left[ \| \epsilon - \epsilon_\theta(\hat{z_t}, t, p, a_i) \|^2_2 \right],
\label{eq:ldm_loss}
\end{equation}

where $\hat{z_t}$ is the noisy latent representation at timestep $t$, $\mathcal{E}(p)$ is the latent representation $\hat{z_0}$ extracted from reference portrait $p$. As the reference portrait is a condition in denoising, different timestep ranges would have different attack performances. Hence, we empirically experiment on $t$ to find the best range, as shown in Fig.~\ref{fig:fig_timesteps}.

$\mathcal{L}_{N}$ modifies the training target of talking-head generation from a synchronized frame to a still portrait. Then we treat $\mathcal{L}_{N}$ as the adversarial loss $\mathcal{L}_{adv}$ in Eq.~\ref{eq:PGD} and optimize the reference image $p$ with PGD for $n$ iterations to acquire the adversarial example $p^n$, as shown in Fig.~\ref{fig:fig_main} (a). It is noted that we adopt gradient descent rather than gradient ascent to optimize the adversarial portrait:

\begin{equation}
p^{n} = \mathcal{P}_{B_{\infty}(p, \delta)} \left[ p^{n-1} - \eta \, \text{sign} \, \nabla_{p^{n-1}} \mathcal{L}_{N}(p^{n-1}) \right]
    \label{eq:PGD_pn}
\end{equation}

where $p^{n-1}$ is the input image in $n$-th iteration, $p^n$ is the output adversarial image. Not only does $p^n$ disrupt synchronization by remaining ``silent'', but it also degrades the video quality of the talking-head generation.

\subsubsection{Silencer-II: Anti-purification}
By optimizing a perturbation using our proposed nullifying loss, and adding it to the portrait image, we can generate an adversarial example that prevents the portrait from being driven by audio.
Unfortunately, existing noise-removal or ``purification'' techniques can easily strip away the noise, undermining the protection effect. To address this, we need to find a more robust noise pattern that resists these purification methods, enhancing the security and effectiveness of the adversarial examples generated with our \textbf{Silencer}.

\begin{figure}
  \centering
  \includegraphics[width=0.45\textwidth]{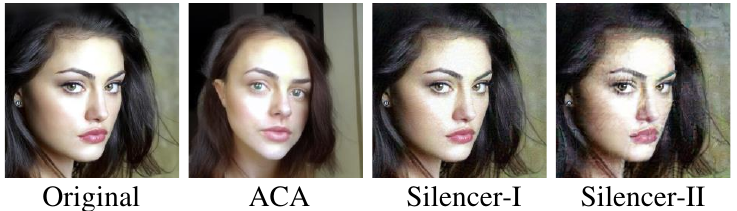}
  \vspace{-0.2cm}
  \caption{\textbf{Visualization of ACA~\cite{chen2023contentbased} and Silencer.} The result of ACA is generated by optimizing latent feature with skip gradients using our nullifying loss.}
  \label{fig:fig_aca} 
  \vspace{-0.4cm}
\end{figure}

\begin{figure*}
  \centering
  \includegraphics[width=1.\textwidth]{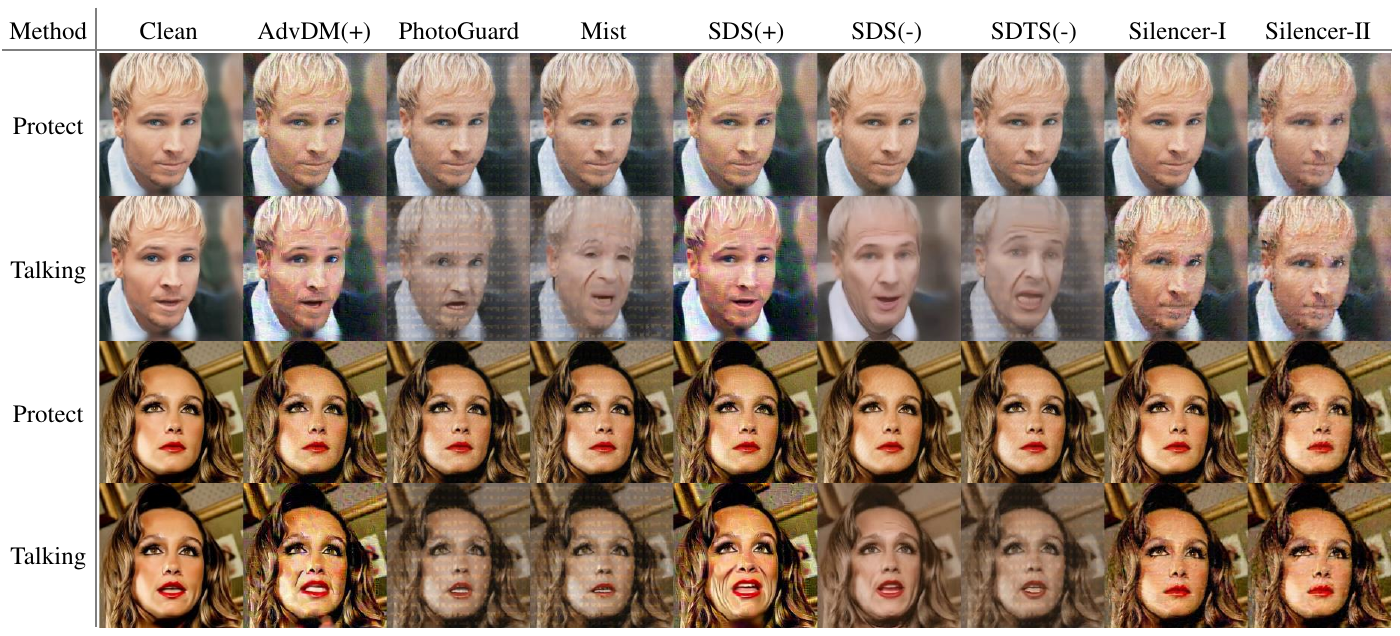}
  \vspace{-0.5cm}
  \caption{\textbf{Qualitative Comparison with Image Protection Methods.} We visualize the protected portraits and their frame driven by audio.}
  \label{fig:fig_quality1} 
  \vspace{-0.3cm}
\end{figure*}

\begin{table}
\setlength{\tabcolsep}{5pt}
\begin{center}
\small
\resizebox{0.48\textwidth}{!}{
\setlength\tabcolsep{6pt}
\renewcommand\arraystretch{1.}
\begin{tabular}{z{60}y{22}|| c | c}
\hline \thickhline
\rowcolor{mygray}
 & & CelebA-HQ  & TalkingHead-1KH 
\\ \cline{3-4} \rowcolor{mygray} 
              \multicolumn{2}{c||}{\multirow{-2}{*}{Method}} & I-PSNR/SSIM$\uparrow$ & I-PSNR/SSIM$\uparrow$ \\ \hline \hline
AdvDM(+)~\cite{liang2023adversarial}\!\!\!&\!\!\!\pub{ICML23}\! &31.15/\textcolor{blue}{0.7605} &31.32/0.7262   \\
AdvDM(-)~\cite{xue2023toward}\!\!\!&\!\!\!\pub{ICLR24}\! &31.02/0.7191 &31.16/0.6807   \\
PhotoGuard~\cite{salman2023raising}\!\!\!&\!\!\!\pub{Arxiv23}\! &29.96/0.7299 &30.20/0.7147  \\
Mist~\cite{liang2023mist}\!\!\!&\!\!\!\pub{Arxiv23}\! &30.06/0.7342 &30.32/0.7190  \\
SDS(+)~\cite{xue2023toward}\!\!\!&\!\!\!\pub{ICLR24}\!  &31.15/\textcolor{red}{0.7688} &31.29/0.7341   \\
SDS(-)~\cite{xue2023toward}\!\!\!&\!\!\!\pub{ICLR24}\! &\textcolor{blue}{31.26}/0.7374 &\textcolor{blue}{31.50}/0.7062   \\
SDTS(-)~\cite{xue2023toward}\!\!\!&\!\!\!\pub{ICLR24}\! &30.42/0.7446 &30.76/0.7307    \\ \hline
\eat{\textbf{Silencer}}{Stage I} &  &\textcolor{red}{31.36}/0.7475 &\textcolor{red}{32.34}/\textcolor{blue}{0.7353}   \\
\eat{\textbf{Silencer}}{Stage II} & &27.23/0.6774 &29.05/\textcolor{red}{0.7590}  \\ 
\hline
\end{tabular}
}
\end{center}
\vspace{-0.5cm}
\caption{{\textbf{Comparison on Image Quality after Protection.} Our Silencer-I achieves the best average image quality with minimal added noise while achieving protection effects.}}
\label{tab:tab_iquality}
\end{table}

ACA~\cite{chen2023contentbased} generates adversarial examples by applying the gradients of adversarial classification loss in the latent space inverted by DDIM~\cite{song2020denoising}. While it can make natural modifications to image content, optimizing the latent vector $z_T$ through the skipped gradients can lead to unpredictable and undesirable changes, compromising the authenticity of the adversarial sample.
These deviations are particularly pronounced in talking-head generation models, leading to significant distortions in facial identity, as shown in Fig.~\ref{fig:fig_aca}.
Existing methods to address this issue, such as applying consistent constraints throughout the inversion process\cite{10716799}, suffer from high memory consumption and are not scalable to high-resolution portrait images.

To address the limitations of existing methods, we propose a new constrain for generating robust adversarial examples with lower computational cost. Our method, illustrated in Fig.\ref{fig:fig_main} (b), leverages LDM and DDIM inversion to optimize the latent representation of the image. Instead of directly adding noise, we optimize the latent features to create perturbations that are resistant to purification techniques. To ensure these perturbations are effective without significantly altering essential facial features, we utilize a constraint during the optimization process. Specifically, we extracted the VAE feature of adversarial samples generated in step I. The encoded features then serve as a constraint, during the optimization of the LDM-based adversarial example. This constrained optimization allows us to balance two objectives: maintaining crucial facial features for recognition, while simultaneously maximizing the robustness of the perturbations against purification defenses. The optimization objective is formulated as follows:

\begin{align}
\mathcal{L}_{AP} &=\lambda_1\mathcal{L}_N-\lambda_2\mathcal{L}_T\\
 &= \lambda_1\mathcal{L}_N + \lambda_2\| \mathcal{E}(p'_0) - \mathcal{E}(p^n) \|_2^2
    \label{eq:PGD_total_loss}
\end{align}

where $p'_0$ is the output of the inverted diffusion model, $p^n$ is the adversarial example generated in stage I, $\lambda_1$ and $\lambda_2$ are the corresponding coefficients. Then we use AdamW~\cite{loshchilov2017decoupled} to optimize the inverted latent representation $p_t$ with $\mathcal{L}_{AP}$. 

We observed that optimizing the entire image with $\mathcal{L}_{AP}$ introduces considerable distortions to the facial region. LDM tends to distort the entire face in the iterative optimization process, aiming to eliminate recognizable features like the mouth and eyes. This results in a damaged face without any specific regions that could be manipulated or driven. Hence, we apply a facial mask to the optimization process that strikes a balance between face clarity and anti-purification protection. Specifically, we optimize the entire image during the initial $s$ iterations. After this point, we restrict optimization to areas outside the masked facial region.




\section{Experiments}
\subsection{Experimental Setup}
\subsubsection{Implementation Details}


The videos are sampled at 25 FPS and the audio sample rate is 16KHz. The reference portraits are resized to 512 $\times$ 512. We utilize Hallo~\cite{xu2024hallo} as the LDM-based talking-head model with the public implementation\footnote{https://github.com/fudan-generative-vision/hallo}.
In the first stage, we adopt PGD with a budget 16/255 to train each portrait for 100 iterations, which is the same as our baselines. In the second stage, we optimize the inverted latent feature for 200 iterations with a learning rate of 0.01.

\paragraph{Baselines and Dataset.} We compare our proposed method with four state-of-the-art privacy protection methods, including AdvDM~\cite{liang2023adversarial}, PhotoGuard~\cite{salman2023raising}, Mist~\cite{liang2023mist} and SDS~\cite{xue2023toward}. To evaluate the performance of protection baselines, we select 50 images from the CelebA-HQ~\cite{karras2018progressive} dataset as the reference images and one audio as the driving signal. To create a more realistic scenario, we utilize CLIP-IQA~\cite{wang2023exploring} to select 50 high-quality video clips, each with a unique identity and associated speech audio, from the widely used TalkingHead-1KH dataset~\cite{wang2021facevid2vid}.

\begin{table*}
\setlength{\tabcolsep}{5pt}
\begin{center}
\small
\resizebox{0.999\textwidth}{!}{
\setlength\tabcolsep{6pt}
\renewcommand\arraystretch{1.}
\begin{tabular}{z{65}y{22}|| c | c c c c c}
\hline \thickhline
\rowcolor{mygray}
 & & Protected & JPEG~\cite{sandoval2023jpeg}  & AdvClean &DiffPure~\cite{nie2022diffusion}     & GrIDPure~\cite{zhao2024can}  
\\ \cline{3-7} \rowcolor{mygray} 
              \multicolumn{2}{c||}{\multirow{-2}{*}{Method}} & I-PSNR/FID & I-PSNR$\downarrow$/FID$\uparrow$ & I-PSNR$\downarrow$/FID$\uparrow$ & I-PSNR$\downarrow$/FID$\uparrow$ & I-PSNR$\downarrow$/FID$\uparrow$ \\ \hline \hline
AdvDM(+)~\cite{liang2023adversarial}\!\!\!&\!\!\!\pub{ICML23}\!  & 31.15/86.58 & 31.35/62.09 & 33.65/49.36 & 28.78/44.83 & 27.81/38.55    \\
AdvDM(-)~\cite{xue2023toward}\!\!\!&\!\!\!\pub{ICLR24}\! & 31.02/66.73 & 31.96/36.97 & 33.29/40.17 & 29.11/42.17 & 27.86/25.48   \\
PhotoGuard~\cite{salman2023raising}\!\!\!&\!\!\!\pub{Arxiv23}\! & 29.96/153.50 & \textcolor{blue}{30.37}/96.98 & \textcolor{blue}{31.37}/\textcolor{blue}{109.87} & \textcolor{blue}{28.36}/\textcolor{blue}{48.43} & \textcolor{blue}{26.32}/\textcolor{blue}{50.04}   \\
Mist~\cite{liang2023mist}\!\!\!&\!\!\!\pub{Arxiv23}\!  & 30.06/156.42 & 30.47/\textcolor{blue}{99.72} & 31.51/108.92 & 28.40/44.14 & 26.35/49.88  \\
SDS(+)~\cite{xue2023toward}\!\!\!&\!\!\!\pub{ICLR24}\!   & 31.15/86.41 & 31.30/60.47 & 33.58/44.80 & 28.72/44.23 & 27.81/38.29    \\
SDS(-)~\cite{xue2023toward}\!\!\!&\!\!\!\pub{ICLR24}\!  & 31.26/70.93 & 32.24/40.69 & 33.31/48.07 & 29.10/39.92 & 27.89/25.90   \\
SDTS(-)~\cite{xue2023toward}\!\!\!&\!\!\!\pub{ICLR24}\!  & 30.42/112.43 & 30.97/72.90 & 31.89/82.19 & 28.53/47.97 & 26.63/39.90     \\ \hline
\eat{\textbf{Silencer}}{Stage I} &  & 31.36/135.77 & 33.38/55.94 & 35.03/61.38 & 29.26/38.93 & 28.16/20.85     \\
\eat{\textbf{Silencer}}{Stage II} & & 27.23/175.21 & \textcolor{red}{27.41}/\textcolor{red}{159.34} & \textcolor{red}{27.95}/\textcolor{red}{135.30} & \textcolor{red}{27.26}/\textcolor{red}{87.22} & \textcolor{red}{25.72}/\textcolor{red}{144.89}    \\ 
\hline
\end{tabular}
}
\end{center}
\vspace{-0.4cm}
\caption{\small{\textbf{Purification Experiments on CelebA-HQ~\cite{karras2018progressive}.} The ``Protected'' is the metrics calculated with protected portraits for reference. Others are calculated with purified portraits. "$\uparrow$": higher is better. "$\downarrow$": lower is better. \textcolor{red}{Red}: the 1st score. \textcolor{blue}{Blue}: the 2nd score.}}

\label{tab:tab_pure}
\end{table*}

\begin{figure*}
  \centering
  \includegraphics[width=1.\textwidth]{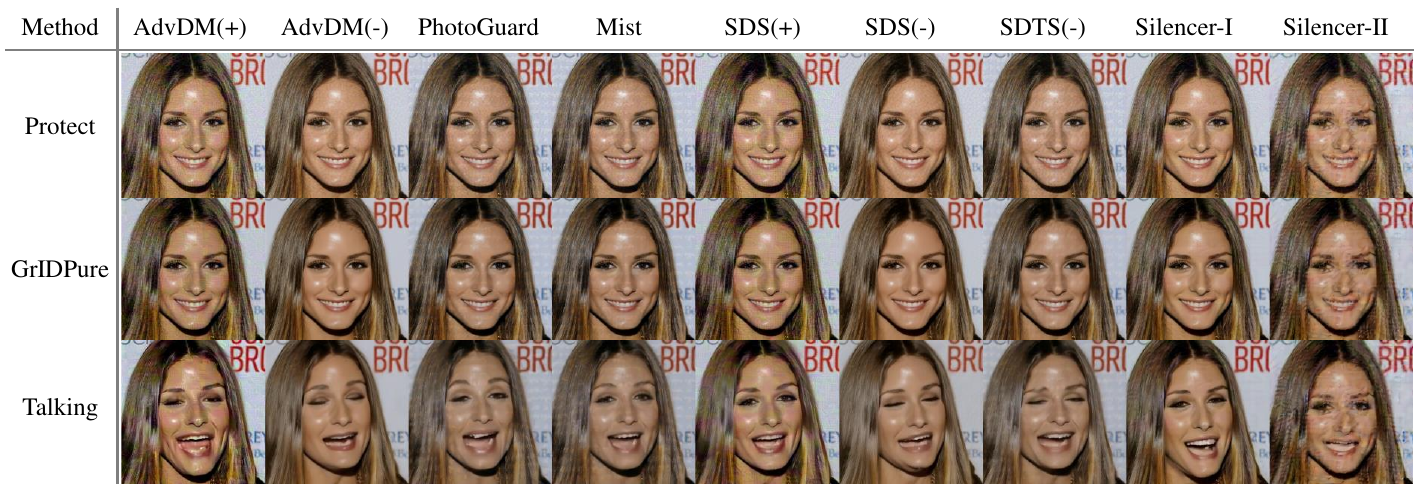}
  \vspace{-0.5cm}
  \caption{\textbf{Visual Comparison in Anti-purification.} The third row is the animated talking frames with the portraits after GrIDPure~\cite{zhao2024can}.}
  \label{fig:fig_quality2} 
  \vspace{-0.1cm}
\end{figure*}





\subsubsection{Metrics}
We assess the quality of synthesized emotional videos with the following metrics:


\noindent\textbf{Image quality.} We utilize Peak Signal-to-Noise Ratio (PSNR), Structural Similarity Index Measure (SSIM), and Fréchet Inception Distance score (FID)~\cite{heusel2017gans} to measure the image quality of synthesized videos. The video metrics, V-PSNR/SSIM, measure PSNR/SSIM specifically on facial regions. In contrast, the image metrics, I-PSNR/SSIM, calculate PSNR and SSIM across the entire image by comparing the original image with the adversarial image. The Fréchet Inception Distance (FID) is a common metric for measuring the fidelity of synthesized videos. It quantifies the distribution distance between videos generated using original and protected portraits.

\noindent\textbf{Audio-visual synchronization.} We evaluate the audio-visual synchronization of the synthesized videos using SyncNet's confidence score~\cite{chung2016out, gan2023efficient}. In addition, the distance between the landmarks of the mouth (M-LMD)~\cite{chen2019hierarchical} is used to indicate speech content consistency.

\subsection{Privacy Protection}
We first compare the effectiveness of our Silencer in privacy protection with other state-of-the-art methods.We randomly selected an audio from TalkingHead-1KH dataset for training all adversarial example and tested the talking-head model with other audios. In CelebA-HQ, all tests used the same audio clip, while in TalkingHead-1KH, each face was tested with its original audio. We treat videos generated by Hallo using the portraits without protection as the ground truth for comparison.

Table~\ref{tab:tab_1} shows that our method achieves the best synchronization protection, with a score of 3.9685 on CelebA-HQ and 2.0017 on TalkingHead-1KH. In stage I of Silencer, our nullifying loss effectively frees reference portraits from audio control during talking-head generation. In stage II, our Silencer continues to yield strong results, further validating the effectiveness of our method. In terms of video quality, our method can only achieve comparable results in the video FID. This is primarily due to our nullifying loss, which aims to ensure the reference portrait remains largely unchanged during the diffusion process. As a result, our method has less impact on the generated video quality compared to others. Table~\ref{tab:tab_iquality} shows that Silencer-I achieves the highest I-PSNR, indicating minimal degradation of the reference portrait's realism. Furthermore, the qualitative comparison in Fig.~\ref{fig:fig_quality1} reveals that, unlike methods that significantly alter facial appearance, our approach  preserves visual consistency while ``silencing'' the talking head. These results underscore the effectiveness of our method in achieving privacy protection from audio control.

\subsection{Anti-Purification Experiments}
To demonstrate the effectiveness of our methods in resisting purification, we conduct purification on the protected portraits. We select the following advanced purification methods to attack: JPEG~\cite{sandoval2023jpeg}, AdvClean, Diff-Pure~\cite{nie2022diffusion} and GrIDPure~\cite{zhao2024can}.We evaluate anti-purification effectiveness using I-PSNR, comparing original and purified images, and FID for talking-head videos, comparing ground truth to videos generated with purified portraits from a CelebA-HQ subset. Table~\ref{tab:tab_pure} shows Silencer-II achieves the best anti-purification performance, with the lowest I-PSNR and highest FID. This is because adversarial noises, generated by Silencer-I and other methods, are optimized in the image space with PGD~\cite{madry2017towards} and can be easily purified. In contrast, Silencer-II optimizes perturbations within the LDM's inverted latent space, resulting in fundamentally different and more robust perturbations. Fig.~\ref{fig:fig_quality2} visually demonstrates this efficacy.
Although our generated perturbations resist complete removal, their structure is altered by purification, preventing a perfectly ``silent'' portrait. Achieving a completely robust perturbation that results in a ``silent'' portrait even after purification remains a challenge. We will explore more robust solutions in future work.

\begin{figure}
  \centering
  \includegraphics[width=.48\textwidth]{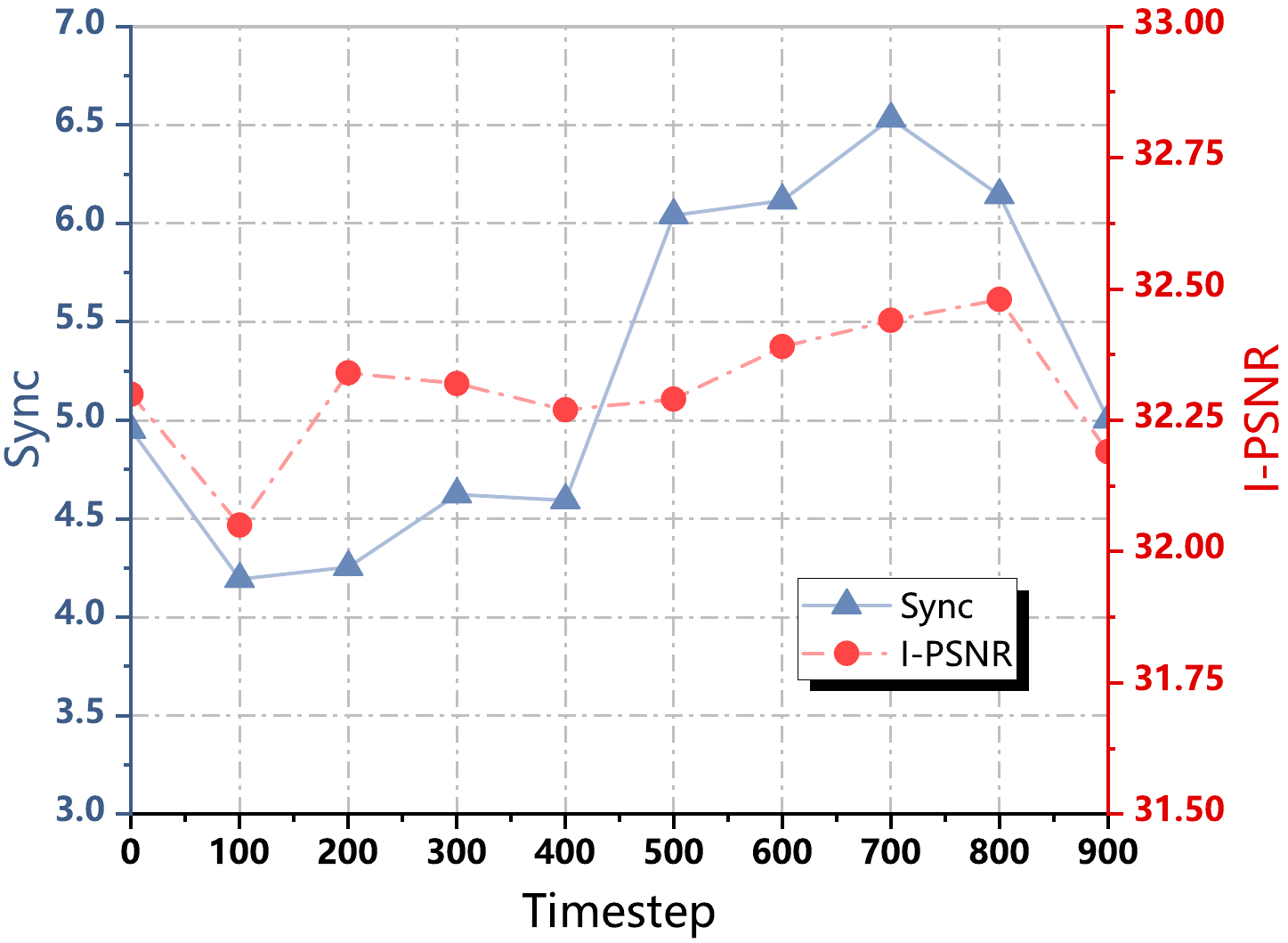}
  \vspace{-0.4cm}
  \caption{\textbf{Ablation Study on Timestep Ranges in Silencer-I.} }
  \label{fig:fig_timesteps} 
\end{figure}

\subsection{Ablation Study}

\paragraph{Ablation Study on Timestep Ranges in Silencer-I.} Since the reference portrait serves as a condition in the denoising process, adjusting the timestep range results in varying levels of attack effectiveness. To identify the optimal timestep ranges, we divided the total of 1000 timesteps into ten equal segments, sampling 100 timesteps from each segment for training. We trained and evaluated our model on the subset of CelebA-HQ, with the results illustrated in Fig.~\ref{fig:fig_timesteps}. Our findings indicate that timesteps within the [200, 300] range achieve a desirable balance: they provide effective privacy protection, with a sync confidence of 4.2556, while maintaining minimal noise, with an I-PSNR of 32.34. Based on these results, we selected the [200, 300] range for timestep sampling in training Silencer-I.

\paragraph{Ablation Study on Each Component.} We conduct an ablation study to evaluate the impact of each component of our Silencer using the CelebA-HQ dataset. As shown in Table~\ref{tab:tab_ablation}, the introduction of our nullifying loss leads to a significant reduction in synchronization confidence compared to previous methods. Additionally, the anti-purification process remains the low synchronization, providing protection against talking-head manipulation. However, this comes at the cost of reduced visual quality. To mitigate this, we optimize the adversarial perturbation with a face mask, which preserves facial structure while achieving effective privacy protection. More experiments can be found in our supplementary material.

\begin{table}
\setlength{\tabcolsep}{5pt}
\begin{center}
\small
\resizebox{0.48\textwidth}{!}{
\setlength\tabcolsep{6pt}
\renewcommand\arraystretch{1.}
\begin{tabular}{c|| c c c c}
\hline \thickhline
\rowcolor{mygray}
 Ablation & SDTS(-) & S-I & S-II (A) & S-II (B) \\ \hline \hline
$\mathcal{L}_{N}$   &  &  \checkmark & \checkmark   & \checkmark   \\
Anti-purify           &  &   & \checkmark  & \checkmark           \\
Mask                &  &   &    & \checkmark            \\ \hline
I-SSIM$\uparrow$    & \textcolor{blue}{0.7446} & \textcolor{red}{0.7475} & 0.6561 & 0.6774 \\ 
FID$\uparrow$       & 89.70  & 124.07 & \textcolor{red}{167.40} & \textcolor{blue}{156.99} \\ 
Sync$\downarrow$    & 6.4003 & 4.0644 & \textcolor{red}{3.4339} & \textcolor{blue}{3.9685}\\ 
M-LMD$\uparrow$   & 2.1024 & 2.2008 & \textcolor{red}{2.3607}& \textcolor{blue}{2.2108}  \\ 
\hline
\end{tabular}
}
\end{center}
\vspace{-0.1cm}
\caption{{\textbf{Ablation Study of Each Component.} Each component contributes to improving privacy protection, thus verifying its effectiveness.}}
\vspace{-0.2cm}
\label{tab:tab_ablation}
\end{table}



\section{Conclusion}
In this paper, we introduce \textbf{Silencer}, a two-stage approach to proactively protect portrait privacy from unauthorized animation in audio-driven talking-head generation. This approach addresses the limitations of prior methods that cannot effectively mitigate talking animation and resist purification. 
The first stage employs a novel nullifying loss to decouple facial movements from audio input, significantly reducing the synchronization of generated talking-head videos. Building upon this, the second stage enhances robustness through an anti-purification process. This process optimizes perturbations within the inverted latent space of an LDM, guided by adversarial examples from the first stage to ensure targeted and effective protection. A strategically applied mask preserves facial integrity during this optimization.
Extensive experiments demonstrate Silencer's superior performance in both preventing unauthorized animation and resisting purification, confirming its effectiveness in protecting portrait privacy. 
This work establishes a new benchmark for proactive privacy protection in LDM-based talking-head generation and we anticipate it will stimulate further research and development in this critical area.

\section*{Acknowledgments}

This work was supported in part by the Key R\&D Program of China (2021ZD0112801), the Key Program of National Natural Science Foundation of China (62436007) and the Natural Science Foundation of Zhejiang Province (LDT23F02023F02).

{
    \small
    \bibliographystyle{ieeenat_fullname}
    \bibliography{main}
}

\newpage

\appendix

\section{More Experiments}
\label{sec:more_exp}

\subsection{More Implementation Details}
The resolution of our input portrait is $512\times 512$. The audio used for training in our experiment is a four-second clip. For testing on CelebA-HQ, the audio length is seven seconds. In the case of TalkingHead-1KH, the audio length varies between three and seven seconds. In our experiment, the DDIM inversion step is set to 20. Due to the limitation of GPU memory, we optimize only the inverted latent feature from the final step. All experiments can be conducted using a single NVIDIA A40 GPU.

\subsection{Evaluating the Transferability of Silencer}
To evaluate the transferability of {\bf Silencer (S-I and S-II)}, we performed a cross-model evaluation. Adversarial noise was optimized on the Hallo model and subsequently tested on other LDM-based talking-head generation models.
Specifically, we randomly selected 20 portraits from the TalkingHead-1KH dataset and generated talking-head videos using the publicly available EchoMimic~\cite{chen2024echomimic} and Hallo2~\cite{cui2024hallo2}. 
As shown in Table~\ref{tab:transferability}, the synchronization values of the generated videos demonstrate that Silencer maintains a significant adversarial effect even when applied to models different from the one used for optimization.
Although \textbf{Silencer} 
is designed as a white-box attack, these results highlight its notable generalization capability across various LDM-based talking-head models.
This cross-model robustness suggests the potential for broader applicability and further validates the effectiveness of our method. A likely explanation for the observed cross-model effectiveness of Silencer is a combination of factors. First, these LDM-based talking-head models share similar architectural designs. Second, and perhaps more crucially, they are all fine-tuned upon Stable Diffusion. This common foundation could introduce common weaknesses or biases that Silencer is able to exploit, even across different models.

\subsection{Efficiency Analysis}
We evaluated the computational efficiency on an NVIDIA A40 GPU. The results, shown in Table~\ref{tab:efficiency}, demonstrate a significant difference in Silencer-I and Silencer-II. Silencer-I exhibits superior efficiency, requiring considerably less computational time compared to Silencer-II. This difference in efficiency stems primarily from the architectural design of Silencer-II. Unlike Silencer-I, Silencer-II incorporates an optimization step within the latent space of an additional LDM. This additional optimization process introduces a substantial computational overhead, increasing the overall time required for Silencer-II to generate adversarial examples. While this optimization contributes to more robust perturbations, it comes at the cost of reduced computational efficiency. Silencer-I, by contrast, avoids this extra optimization step, leading to a more streamlined and faster process. While Silencer-I takes 64 seconds per image, its runtime is comparable to other methods like AdvDM(+) and Mist (59 seconds). This makes Silencer-I a more practical choice in scenarios where computational resources are limited or where rapid generation of adversarial examples is critical. Notably, SDS(-) demonstrate significantly faster runtimes, due to skipping the UNet portion of the gradient calculation. However, whether such an optimization can be effectively and reliably applied within an LDM-based talking-head network to improve efficiency remains an open challenge for future research.

\begin{table}
\setlength{\tabcolsep}{5pt}
\begin{center}
\small
\resizebox{0.48\textwidth}{!}{
\setlength\tabcolsep{2pt}
\renewcommand\arraystretch{1.}
\begin{tabular}{c|| c c c c c c}
\hline \thickhline
\rowcolor{mygray}
 Method & GT & AdvDM(+) & Mist & SDST(-) & \textbf{S-I} & \textbf{S-II} \\ \hline \hline
EchoMimic~\cite{chen2024echomimic} & 4.0365 & 1.8252 & 1.7839 & 2.2228 & \textcolor{blue}{1.4601} &  \textcolor{red}{0.9973}   \\ 
Hallo2~\cite{cui2024hallo2} & 5.6661 & 3.2136 & 3.0679 & 3.9238 & \textcolor{red}{1.5952} &  \textcolor{blue}{2.0783}   \\ 
\hline
\end{tabular}
}
\end{center}
\caption{{\textbf{Evaluating the Transferability of Silencer.}}  Synchronization scores demonstrating cross-model transferability of Silencer (S-I and S-II). Videos were generated by EchoMimic~\cite{chen2024echomimic} and Hallo2~\cite{cui2024hallo2} using original (GT) and adversarial inputs. Lower scores signify greater disruption. Despite being optimized on Hallo, Silencer significantly impacts both models.}
\label{tab:transferability}
\end{table}

\begin{table}
\setlength{\tabcolsep}{1pt}
\begin{center}
\small
\resizebox{0.5\textwidth}{!}{
\setlength\tabcolsep{2pt}
\renewcommand\arraystretch{1.}
\begin{tabular}{c|| c c c c c c c}
\hline \thickhline
\rowcolor{mygray}
  & AdvDM(+) & PhotoGuard & Mist & SDS(-) & SDST(-) & S-I & S-II \\ \hline \hline
 time & 59 & 34 & 59 & 22 & 40 & 64 & 241 \\ 
\hline
\end{tabular}
}
\end{center}
\caption{{\textbf{Efficiency Analysis.} Average time (seconds/image) required for different protection methods.}}
\label{tab:efficiency}
\end{table}

\begin{table}
\setlength{\tabcolsep}{1pt}
\begin{center}
\small
\resizebox{0.5\textwidth}{!}{
\setlength\tabcolsep{3pt}
\renewcommand\arraystretch{1.}
\begin{tabular}{c|| c c c }
\hline \thickhline
\rowcolor{mygray}
 DiffPure timesteps & 50 & 100 & 150 \\ \hline \hline
 Silencer-I & 30.65/0.2606  & 29.26/0.2540  & 28.13/0.2691  \\ 
 Silencer-II & \textcolor{red}{27.80}/\textcolor{red}{0.4057} & \textcolor{red}{27.26}/\textcolor{red}{0.3909}  & \textcolor{red}{26.82}/\textcolor{red}{0.3504}  \\ 
\hline
\end{tabular}
}
\end{center}
\caption{{\textbf{Ablation on Timesteps of DiffPure~\cite{nie2022diffusion}.} We present I-PSNR/LPIPS scores for Silencer-I and Silencer-II after applying DiffPure with varying timesteps. Red values highlight greater robustness.}}
\label{tab:timestep_DiffPure}
\end{table}

\begin{table}
\setlength{\tabcolsep}{1pt}
\begin{center}
\small
\resizebox{0.5\textwidth}{!}{
\setlength\tabcolsep{3pt}
\renewcommand\arraystretch{1.}
\begin{tabular}{c|| c c c }
\hline \thickhline
\rowcolor{mygray}
 GrIDPure timesteps & 5 & 10 & 15 \\ \hline \hline
 Silencer-I & 28.35/0.1672 & 28.16/0.1698  &  27.93/0.2016 \\ 
 Silencer-II & \textcolor{red}{25.81}/\textcolor{red}{0.3451} & \textcolor{red}{25.72}/\textcolor{red}{0.3511}  & \textcolor{red}{25.59}/\textcolor{red}{0.3610} \\ 
\hline
\end{tabular}
}
\end{center}
\caption{{\textbf{Ablation on Timesteps of GrIDPure~\cite{zhao2024can}.}} We present I-PSNR/LPIPS scores for Silencer-I and Silencer-II after applying GrIDPure purification. GrIDPure was run for 20 iterations with initial timesteps of 5, 10, and 15. Red values highlight greater robustness.}
\label{tab:timestep_GrIDPure}
\end{table}

\begin{table}
\setlength{\tabcolsep}{1pt}
\begin{center}
\small
\resizebox{0.5\textwidth}{!}{
\setlength\tabcolsep{16pt}
\renewcommand\arraystretch{1.}
\begin{tabular}{c|| c c}
\hline \thickhline
\rowcolor{mygray}
   & DiffAudio & SameAudio  \\ \hline \hline
 Silencer-II & 3.9685 & 2.4926 \\ 
 Ground Truth & 6.4041 & 5.7509 \\
\hline
\end{tabular}
}
\end{center}
\caption{{\textbf{Impact of Audio Consistency on Silencer-II while Training and Testing with CelebA-HQ.} "DiffAudio" denotes using different audio for training and testing, while "SameAudio" uses the same audio. Lower Sync value is better.}}
\label{tab:ab_sameaudio}
\end{table}

\begin{table}
\setlength{\tabcolsep}{5pt}
\begin{center}
\small
\resizebox{0.5\textwidth}{!}{
\setlength\tabcolsep{6pt}
\renewcommand\arraystretch{1.}
\begin{tabular}{c|| c c c c}
\hline \thickhline
\rowcolor{mygray}
 $l_{inf}$ & V-PSNR/SSIM$\downarrow$ &FID$\uparrow$ & Sync$\downarrow$ & M-LMD$\uparrow$ \\ \hline \hline
 8/255  & 19.59/0.5768 & 78.78 & 4.8368 & 2.0444  \\ 
 16/255 & \textcolor{red}{19.02}/\textcolor{red}{0.5104} & \textcolor{red}{124.07} & \textcolor{red}{4.0644} & \textcolor{red}{2.2008}  \\ 
\hline
\end{tabular}
}
\end{center}
\caption{{\textbf{Ablation Study of $l_{inf}$ Perturbation Budgets in Silencer-I on CelebA-HQ.}}}
\label{tab:adv_budget}
\end{table}

\begin{table}
\setlength{\tabcolsep}{5pt}
\begin{center}
\small
\resizebox{0.5\textwidth}{!}{
\setlength\tabcolsep{3pt}
\renewcommand\arraystretch{1.}
\begin{tabular}{c|| c c c c}
\hline \thickhline
\rowcolor{mygray}
 Inverted Timesteps & V-PSNR/SSIM$\downarrow$ &FID$\uparrow$ & Sync$\downarrow$ & M-LMD$\uparrow$ \\ \hline \hline
 the last one & \textcolor{red}{19.01}/\textcolor{red}{0.5111} & \textcolor{red}{156.99} & \textcolor{red}{3.9685} & \textcolor{red}{2.2108} \\ 
 the last two & 19.30/0.5402 & 111.99 & 4.4579 & 2.1731  \\ 
\hline
\end{tabular}
}
\end{center}
\caption{{\textbf{Ablation Study of Inverted Timesteps in Silencer-II on CelebA-HQ.}}}
\label{tab:inverted_timesteps}
\end{table}

\subsection{More Ablation Study}

\paragraph{Ablation Study on Timesteps in Purification Methods.} Our anti-purification experiments are conducted using the publicly available implementation\footnote{https://github.com/zhengyuezhao/gridpure}. For DiffPure, we set the diffusion timestep to 100, while for GrIDPure, we use a timestep of 10 with 20 iterations. 
We conduct the ablation experiments on different settings of diffusion-based purification.
Table~\ref{tab:timestep_DiffPure} and Table~\ref{tab:timestep_GrIDPure} illustrate the effectiveness of Silencer-I and Silencer-II against image purification techniques, specifically DiffPure and GrIDPure, across different timesteps. The tables compare I-PSNR and LPIPS scores for images processed by both Silencer versions. While larger timesteps in these purification methods improve the smoothness of the resulting images, they fail to completely remove the perturbations introduced by Silencer-II. This highlights the robustness of our approach.

\begin{figure}
  \centering
  \includegraphics[width=.48\textwidth]{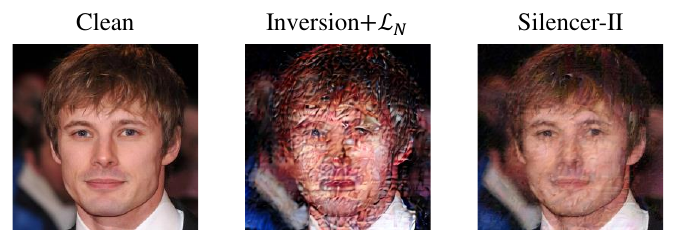}
  \vspace{-0.4cm}
  \caption{\textbf{Ablation Study on $\mathcal{L}_T$ in Silencer-II.} Without the assistance of $\mathcal{L}_T$, the generated perturbation becomes highly noticeable, significantly compromising the facial identity.}
  \label{fig:fig_ab_inversion_ln} 
\end{figure}

\begin{figure}
  \centering
  \includegraphics[width=.49\textwidth]{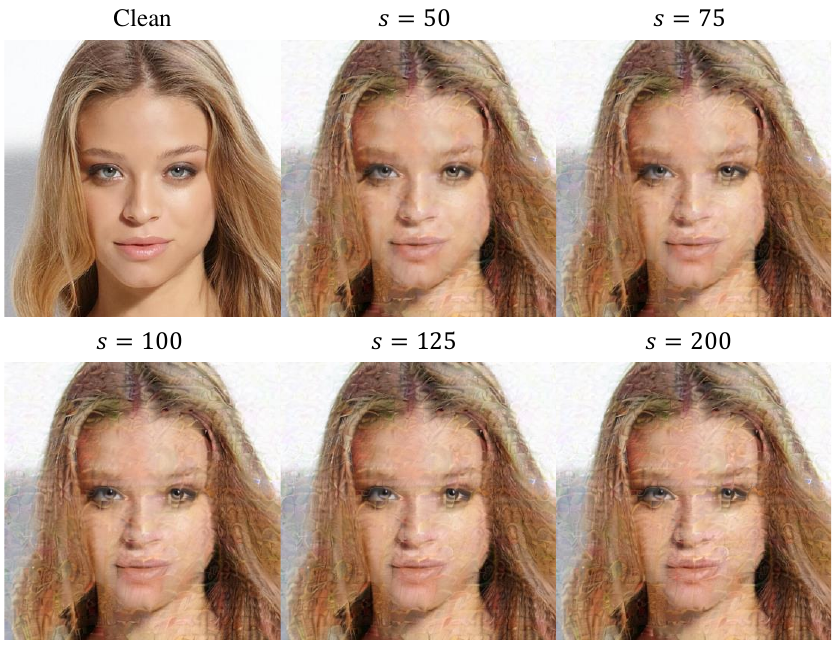}
  \caption{\textbf{Visualization Results with Different Iteration $s$.} The quality of the portrait decreases with the growth of $s$.}
  \label{fig:fig_vis_iter_s} 
\end{figure}

\begin{table}
\setlength{\tabcolsep}{5pt}
\begin{center}
\small
\resizebox{0.48\textwidth}{!}{
\setlength\tabcolsep{6pt}
\renewcommand\arraystretch{1.}
\begin{tabular}{c|| c c c c c}
\hline \thickhline
\rowcolor{mygray}
 $s$ & 50 & 75 & 100 & 125 & 200 \\ \hline \hline
I-SSIM$\uparrow$   & \textcolor{red}{0.7125} & \textcolor{blue}{0.6998} & 0.6918 & 0.6844 & 0.6704    \\ 
FID$\uparrow$      & 136.88 & 166.10 & \textcolor{blue}{173.18} & 171.08 & \textcolor{red}{193.46}    \\ 
Sync$\downarrow$   & 5.5725 & 5.0413 & \textcolor{red}{4.0602} & 4.1832 & \textcolor{blue}{4.0791}    \\ 
M-LMD$\uparrow$    & 1.8559 & 2.1371 & 2.2053 & \textcolor{red}{2.3748} & \textcolor{blue}{2.3563}    \\ 
\hline
\end{tabular}
}
\end{center}
\caption{{\textbf{Ablation Study on the Initial Iteration $s$ without Mask.} Larger iterations without the face mask lead to better protection performance with lower image quality.}}
\label{tab:ab_iter_s}
\end{table}

\begin{figure*}
  \centering
  \includegraphics[width=.99\textwidth]{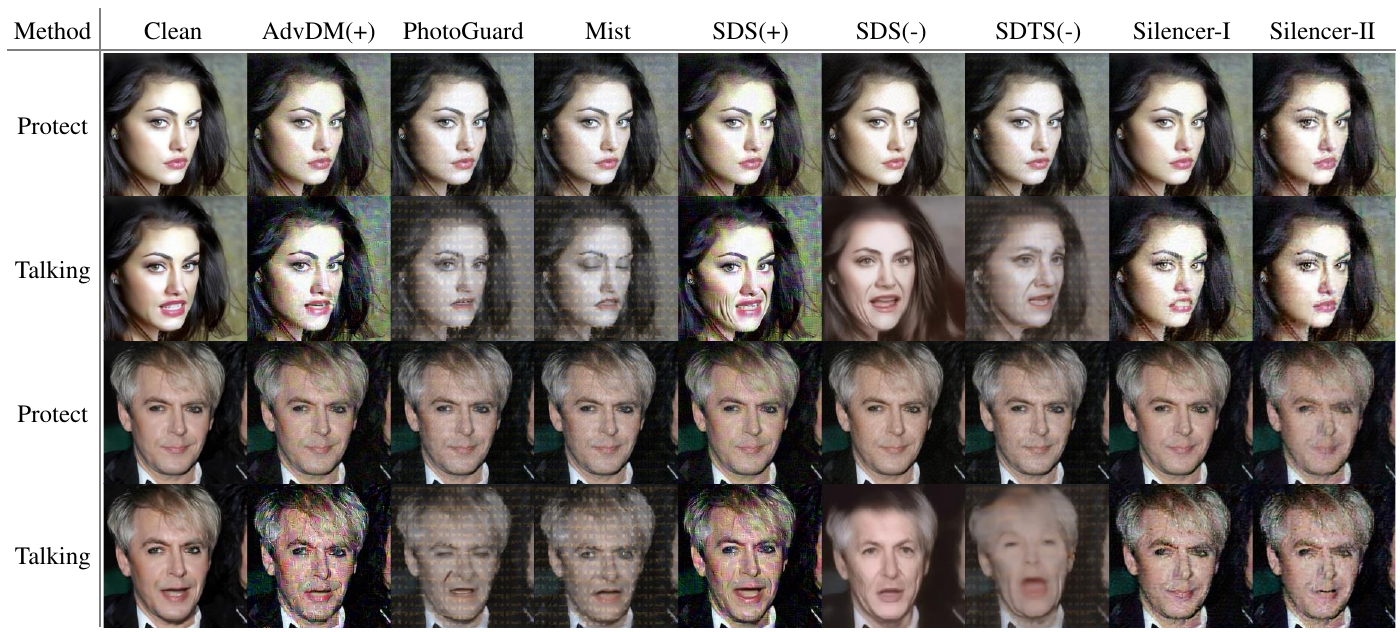}
  \caption{\textbf{Additional Visualization Comparison with Image Protection Methods in CelebA-HQ~\cite{karras2018progressive}.} }
  \label{fig:fig_more_vis_celebahq} 
\end{figure*}

\begin{figure*}[h]
  \centering
  \includegraphics[width=.99\textwidth]{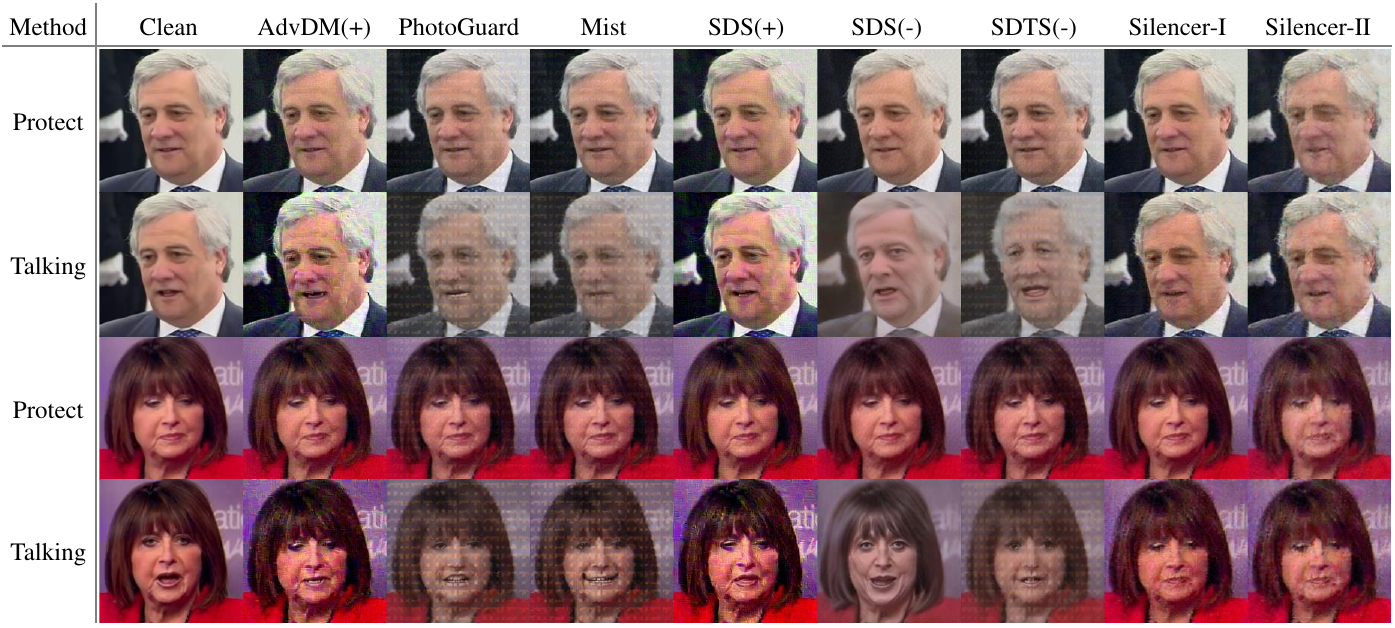}
  \caption{\textbf{Additional Visualization Comparison with Image Protection Methods in TalkingHead-1KH~\cite{wang2021facevid2vid}.} }
  \label{fig:fig_more_vis_th1kh} 
\end{figure*}

\paragraph{Ablation Study on Audio and Portrait in the Training and Testing of CelebA-HQ.}
For audio, We investigated the effect of using the same versus different audio inputs during the training and testing phases. This tests whether Silencer is overly sensitive to specific audio characteristics or if it can generalize to unseen audio. As shown in Table~\ref{tab:ab_sameaudio}, both scenarios resulted in a reduction of the synchronization value compared to the ground truth. The decrease in synchronization demonstrates that Silencer effectively disrupts synchronization regardless of whether the audio input is consistent between training and testing. This finding highlights the robustness of the Silencer method to variations in audio input, suggesting that it is not overfitting to specific audio features.

For the starting portrait, we conducted
experiments on 50 different portraits of CelebA-HQ in Table 1. The average sync value is 3.9685 and the standard deviation is 1.5607. Our findings indicate that the effectiveness of adversarial perturbations varies across different facial identities, suggesting variations in inherent robustness. We intend to investigate the factors contributing to this variability in future research.

\paragraph{Ablation Study on Perturbation Budget in Silencer-I.} To understand the influence of the perturbation budget on the effectiveness of Silencer-I, we conducted an ablation study on the CelebA-HQ dataset.  Specifically, we investigated the performance of Silencer-I under constrained $l_{inf}$ perturbation budgets. The $l_{inf}$ limits the maximum change allowed for any single pixel value in the input image. A smaller budget implies a more subtle, less perceptible adversarial perturbation.
As shown in Table~\ref{tab:adv_budget}, we evaluated Silencer-I with two different $l_{inf}$ budget: 8/255 and 16/255. The results demonstrate that decreasing the perturbation budget leads to a reduction in Silencer-I's performance. This is because a smaller budget restricts the degree to which Silencer-I can modify the input image to disrupt synchronization.
However, even with a stricter budget, Silencer-I still achieves a notable level of protection performance compared with existing methods in Table 1. This suggests that Silencer-I is more effective, achieving considerable protection with fewer changes to the input portrait.

\paragraph{Ablation Study on Inverted Timesteps in Silencer-II.} We conducted an ablation study on the inverted latent space timesteps used in Silencer-II. Due to memory constraints, we investigated the impact of optimizing the latent feature for the final timestep versus optimizing for the final two timesteps specifically in the context of DDIM inversion. As shown in Table~\ref{tab:inverted_timesteps}, optimizing the latent feature at only the final timestep yielded superior performance while consuming fewer resources compared to optimizing the last two steps. Consequently, we opted for the single-timestep optimization strategy. Further exploration is needed to improve the efficiency and effectiveness of latent feature optimization, addressing potential vulnerabilities to purification methods.

\paragraph{Ablation Study on $\mathcal{L}_T$ in Silencer-II.} We perform an ablation study to evaluate the effectiveness of $\mathcal{L}_T$ in optimizing the inverted latent representation. As shown in Fig.~\ref{fig:fig_ab_inversion_ln}, while the nullifying loss $\mathcal{L}_N$ still produces disturbed results, it achieves this by distorting the portrait, compromising the output's quality and identification. It is mainly because the talking-head model fails to operate effectively when it cannot detect a face, rendering it unable to function as intended. This highlights the necessity of exploring optimized solutions that protect privacy without sacrificing visual integrity. With the assistance of $\mathcal{L}_T$, we can effectively reduce noise in the facial region while achieving our intended objectives. This approach strikes a balance between minimizing distortions and achieving the desired outcomes, enhancing the overall effectiveness of Silencer.

\paragraph{Ablation Study on the Initial Iteration $s$ without Mask in Silencer-II.} To prevent facial blurring, we incorporate a face mask during the training process of Silencer-II. We begin by training the entire image without a mask for $s$ iterations. Subsequently, a face mask is applied to exclude the facial region from further optimization. To verify the effect of $s$, we conduct an ablation study on a subset of CelebA-HQ, as shown in Fig.~\ref{fig:fig_vis_iter_s} and Table~\ref{tab:ab_iter_s}. The results indicate that as the number of iterations $s$ increases, face quality deteriorates while protection performance improves.  Therefore, we set $s=100$ in our main experiments as it offers a balanced trade-off between maintaining facial clarity and achieving effective protection.

\subsection{Additional Visual Results}
Additional qualitative comparisons are presented in Fig.~\ref{fig:fig_more_vis_celebahq} and Fig.~\ref{fig:fig_more_vis_th1kh}. These figures illustrate that our Silencer consistently achieves superior protection performance across various datasets. These video results can be found in our supplementary video.


\end{document}